\documentclass[aps,twocolumn,showpacs,superscriptaddress,amssymb]{revtex4}
\date{\today}
\usepackage{graphicx}
\usepackage{bm}
\begin{document}

\title{Shell-model study of mirror nuclei with modern
charge-dependent \textit{NN} potential}

\author{C. Qi}
%\email{cqi@pku.edu.cn}
\affiliation{School of Physics and MOE Laboratory of Heavy Ion
Physics, Peking University, Beijing 100871, China}
\author{F.R. Xu}
\email{frxu@pku.edu.cn} \affiliation{School of Physics and MOE
Laboratory of Heavy Ion Physics, Peking University, Beijing 100871,
China} \affiliation{Institute of Theoretical Physics, Chinese
Academy of Sciences, Beijing 100080, China} \affiliation{Center for
Theoretical Nuclear Physics, National Laboratory for Heavy Ion
Physics, Lanzhou 730000, China}

\begin{abstract}
The properties of $T=1/2$ mirror nuclei in the $fp$ shell have been
studied with a microscopic residual interaction. The
isospin-nonconserving interaction is derived from a high-precision
charge-dependent Bonn \textit{NN} potential using the folded-diagram
renormalization method. The level structures of the nuclei are
calculated, obtaining excellent agreements with experimental
observations till the $0f_{7/2}$ band termination. The role played
by isospin symmetry breaking on ground-state displacement energies
and mirror energy differences is discussed, which may help to
explain the long-standing Nolen-Schiffer anomaly. Electromagnetic
and weak transition properties are presented, with discussions on
the asymmetry in analogous transitions.
\end{abstract}

\pacs{21.30.Fe, 21.60.Cs, 23.20.Lv, 27.40.+z}

\maketitle

\section{Introduction}
The investigation of mirror nuclei along the $N=Z$ line is of
significant interest since it addresses directly the isospin
symmetry problem in nuclear many-fermion systems. Isospin symmetry
is approximate due to the charge dependence in strong force and the
Coulomb force between protons. Direct evidence of isospin symmetry
breaking (ISB) can be deduced from ground-state displacement
energies (MDE) and excitation-energy differences between analogous
states (MED) in mirror nuclei. In the past decade, numerous
experimental and theoretical efforts have been devoted to mirror
pairs in the lower $fp$ shell and extends the knowledge of MED
evolution patterns to high-spin states
\cite{Zuker02,Martinezpinedo97,Brandolini05,Poves01,
Bentley00,Bentley06,Bentley98,Oleary97,Tonev02,William03}.

The MDE range from a few to tens of MeV, with the dominant origin in
the Coulomb field \cite{Shlomo78}. However, if only the Coulomb
effect is considered, a persistent inaccuracy exists between the
theoretical results of MDE and corresponding experimental
observations \cite{ns}. This long-standing problem is referred to as
Nolen-Schiffer anomaly \cite{ns}, revealing the necessity to
introduce charge symmetry breaking (CSB) in strong force in
depicting nuclear properties
\cite{Miller90,Suzuki92,Brown00,Machleidt01,Muther99,Tsushima99}.
Contributions from the Coulomb field and CSB in strong force to MED
has been shown theoretically, e.g., in Ref. \cite{Zuker02}, finding
that the later is at least as important as the former.

The purpose of this work is to study the structures and decay
properties of $T=1/2$ mirror nuclei in the lower \emph{fp} shell by
shell-model diagonalization method and to investigate the effects of
ISB. We employ an isospin-nonconserving effective Hamiltonian
\cite{interaction} derived microscopically from a high-precision
version of the charge-dependent Bonn (CD-Bonn) nucleon-nucleon
(\textit{NN}) potential \cite{Machleidt01} using the folded-diagram
renormalization method \cite{Kuo90,Jensen95}. The charge dependence
of the \textit{NN} interaction is retained, enabling to quantify its
effect on nuclear properties exactly with full model-space
diagonalizations. In our previous work, the effective Hamiltonian
has been used to study the isospin structures of odd-odd $N=Z$
nuclei in the lower \textit{fp} shell \cite{interaction}.

The essence of the shell model lies in that the true eigen energies
and wave functions of the original many-body Hamiltonian can be
constructed by diagonalizing an effective Hamiltonian in a
constrained finite model space \cite{Towner,Kuo90}. The model-space
dependent effective Hamiltonian can be written as
\begin{equation}\label{ham}
H_{\text{eff}}=H_{\text{o}}'+v_{\text{eff}},
\end{equation}
where
$H_{\text{o}}'=\sum_{\alpha}\varepsilon_{\alpha}'\text{a}^{\dag}_{\alpha}
\text{a}_{\alpha}$ is the effective one-body Hamiltonian with
$\varepsilon_{\alpha}'$ being single-particle energies (SPE). In
general calculations, the effective interaction $v_{\text{eff}}$ is
expressed as two-body matrix elements (TBME) in harmonic oscillator
(HO) basis. The effective interaction can be decomposed into two
parts. The monopole part of the interaction, in combination with the
SPE, gives out the bulk properties of the nuclei. The multipole part
of the effective interaction accounts for the configuration-mixing
which is essential for modern shell-model calculations
\cite{Caurier05}.

The effective interaction is directly related to the underlying
\textit{NN} potential. Due to the unperturbative nature of the QCD
at low energies, most of the knowledge concerning the \textit{NN}
force is from the measurements of nucleon-nucleon and
nucleon-deuteron scattering properties. The scattering behavior can
be well approximated by one-boson-exchange (OBE) potential which has
been commonly employed in modern realistic \textit{NN} forces
\cite{Machleidt01}. The bare \textit{NN} potential, however, is
unsuitable for direct applications in nuclear systems due to the
strong repulsive core. In early practices, the Brueckner G reaction
matrix is introduced to evaluate the effective interaction for a
chosen model space, as done in the derivation of the famous
Kuo-Brown interaction \cite{Kuo68}. The G matrix takes into account
the short-range repulsive behavior of the \textit{NN} potential and
satisfies
\begin{equation}
\langle\Psi| G |\Psi\rangle=\langle\Psi|V|\Psi\rangle,
\end{equation}
where $V$ is the bare $NN$ potential and $|\Psi\rangle$ the
correlated wave function.

Effective interaction with the G matrix may lead to bad behavior
with increasing particle numbers. The folded-diagram renormalization
method is proposed to include the core-polarization effect from the
configuration-mixing outside the model space. Folded and non-folded
diagrams are introduced in evaluating the time-evolution operator in
time-dependent perturbation theory in which both contributions from
the model space and the excluded space are considered. The two kinds
of diagrams can be evaluated from the G matrix.

In 1990s, various high-precision phenomenological OBE potentials
(e.g., AV18 of the Argonne group, Nijm I $\&$ II of the Nigmegen
group and CD-Bonn of the Bonn group) have been proposed by fitting
the huge amount of neutron-proton and proton-proton scattering data
available \cite{Machleidt01}, with reasonable descriptions of the
charge dependence in strong force. Both charge independence and
charge symmetry are related to the symmetric properties of the $NN$
force under rotations in isospin space. The breaking of charge
symmetry and charge independence are mainly due to the mass
splitting of the nucleons (m$_{\text{proton}}\neq$
m$_{\text{neutron}}$) and pions (m$_{\pi^{\pm}}\neq$ m$_{\pi^0}$).
CSB is a special case of CIB, referring to the difference between
the proton-proton and neutron-neutron interactions. CIB means that
all interactions in isospin $T=1$ state, the proton-proton,
neutron-neutron and neutron-proton interactions, are different,
after electromagnetic effects have been removed.

All the OBE potentials have similar low-momentum behavior. In the
present work, we use a new high-precision CD-Bonn potential.
Detailed descriptions on the CD-Bonn potential can be found in Ref.
\cite{Machleidt01}. In the CD-Bonn potential, both CSB and CIB are
embedded in all partial waves with angular momentum $J\leq4$. The
off-shell behavior of the nonlocal covariant Feynman amplitudes used
in the potential can lead to larger binding energies for nuclear
many-body systems, which can help to improve the effective
interaction's performance around the $N=28$ shell closure
\cite{KB3}.

\section{Level structures}

The level structures of $T=1/2$ mirror nuclei in the lower $fp$
shell are calculated with the residual interaction described above
(denoted as CD-Bonn). For comparison, the isospin-conserving KB3
interaction \cite{KB3} is also used. KB3 is a revised version of the
Kuo-Brown interaction with monopole centroids modified to improve
its performance around $^{56}$Ni \cite{KB3}. The effective
Hamiltonians are diagonalized with the shell model code OXBASH
\cite{Oxbash}. Calculations are performed in the $fp$ major shell.
Specific center-of-mass corrections due to the effects of spurious
states can be avoided \cite{Dean04}. Earlier theoretical efforts on
the nuclei can be find, e.g., in Refs.
\cite{Brandolini05,Poves01,Martinezpinedo97}.

%%%%%%%%%%%%%%%%%%% fig 1 & 2 %%%%%%%%%%%%%%%%%%%%
\begin{figure}
\includegraphics[scale=0.41]{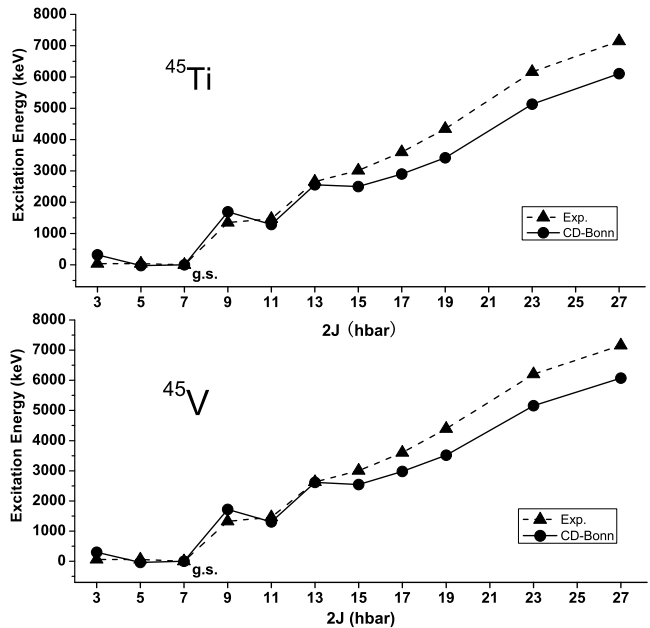}
\caption{\label{fig1}Experimental and calculated excitation energies
for yrast bands in $^{45}$Ti and $^{45}$V. Experimental results are
taken from Ref. \cite{Bentley06}}
\end{figure}

\begin{figure}
\includegraphics[scale=0.53]{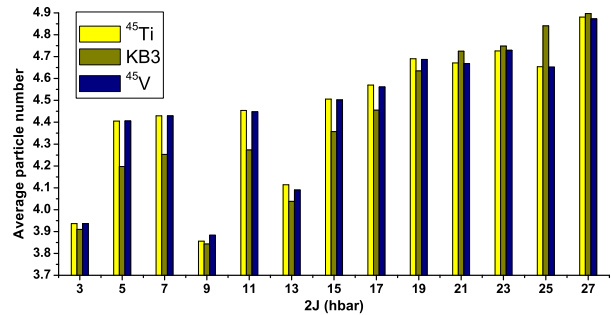}
\caption{\label{fig2} Average number of particles in the $0f_{7/2}$
sub-shell for yrast states in $^{45}$Ti and $^{45}$V. Columns
denoted by $^{45}$Ti and $^{45}$V are results calculated with the
CD-Bonn interaction.}
\end{figure}
%%%%%%%%%%%%%%%%%%%%%%%%%%%%%%%%%%%%%%%%%%%%%%%%%%%%%%%

Fig. \ref{fig1} shows the calculated excitation energies of yrast
bands in $^{45}$V and $^{45}$Ti, together with experimental
observations \cite{Bentley06}. Agreements between calculations and
experiments are satisfactory. Experiments have identified three
low-lying states with $J^{\pi}=7/2^-,~5/2^-$ and $3/2^-$. In the
framework of the deformed Nilsson model, the structure can be
interpreted as the splitting of the $0f_{7/2}$ harmonic oscillator
orbit into $[321]3/2^-,~[312]5/2^-$ and $[303]7/2^-$ orbits at low
excitation energies. The existence of these nearly-degenerate states
gives evidence for nuclear deformation. In spherical shell model,
deformation is described as configuration-mixing contribution from
upper sub-shells. The configuration of $3/2^-$ state can be
approximately described as the excitation of one particle out of the
$0f_{7/2}$ orbit, as shown in Fig. \ref{fig2}, in which the
$0f_{7/2}$ occupancies for $^{45}$V and $^{45}$Ti are plotted as a
function of spin.

%%%%%%%%%%%%%%%%% fig 3 , 4 %%%%%%%%%%%%%%%%%%%%%%%%%%%%%%%%
\begin{figure}
\includegraphics[scale=0.38]{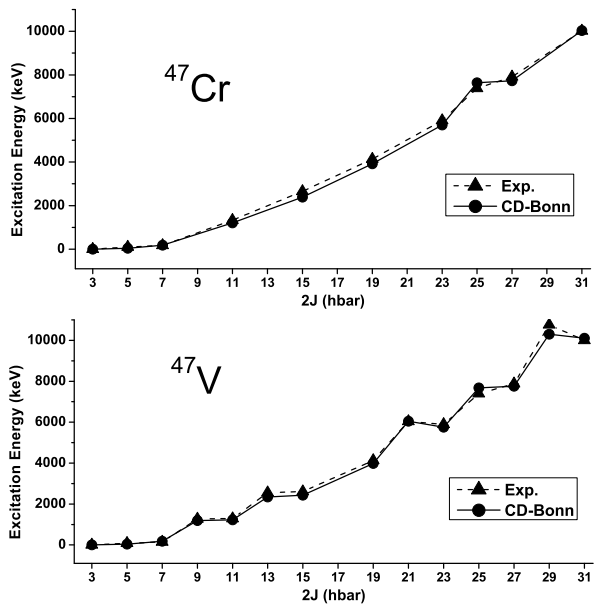}
\caption{\label{fig3}Same as Fig. \ref{fig1} but for $^{47}$V and
$^{47}$Cr. Experimental results are taken from Ref.
\cite{Bentley98,Tonev02}}
\end{figure}

\begin{figure}
\includegraphics[width=0.5\textwidth]{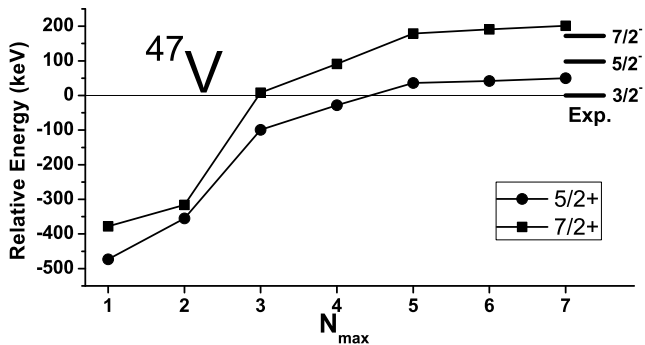}
\caption{\label{fig4} The sequence of the ground-state triplet under
different model truncations. $\text{N}_{\text{max}}$ denotes the
maximal number of particles being excited to the upper $fp$ shell.}
\end{figure}

%%%%%%%%%%%%%%%%%%%%%%% fig 5 %%%%%%%%%%%%%%%%%%%%%%%%
\begin{figure}[htdp]
\includegraphics[width=0.5\textwidth]{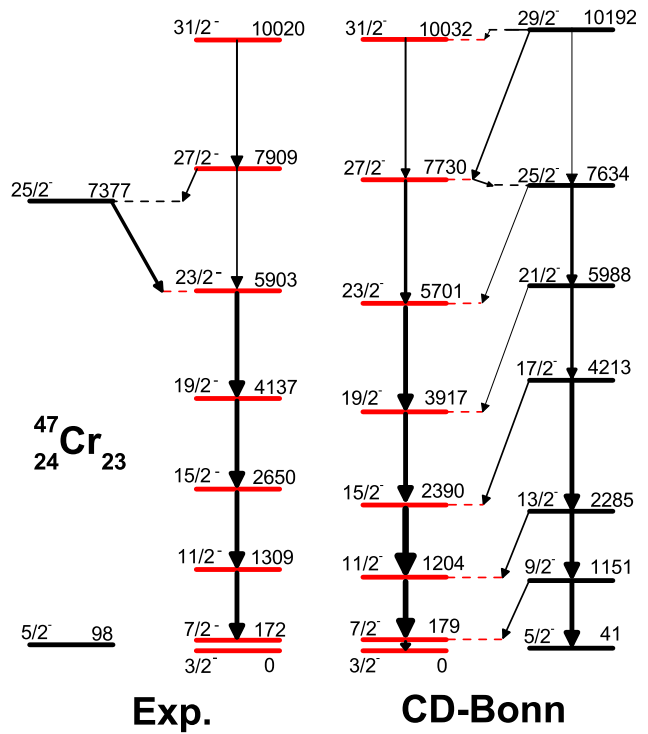}
\caption{\label{fig5}Experimental and proposed yrast band for
$^{47}$Cr.}
\end{figure}

For nuclei with $A\geq 47$, excellent agreements between
calculations and experiments are obtained. Theoretical and
experimental results for mirror pair $^{47}$V and $^{47}$Cr are
shown in Fig. \ref{fig3}. The nucleus $^{47}$V has been recognized
as a $K=3/2$ rotor. Our calculations show that the contribution from
the upper $fp$ shell plays an important role in giving the correct
positions of the $3/2^-$ ground state and other low-lying states. In
Fig. \ref{fig4} we plotted the evolution of the relative positions
of the ground-state triplet as a function of the maximal number of
particles ($\text{N}_{\text{max}}$) being excited out of the
$0f_{7/2}$ sub-shell. To give the correct $3/2^-$ ground state,
$\text{N}_{\text{max}}$ must be larger than four to account for the
configuration mixing. This configuration corresponds to a prolate
deformation. Also interesting is the nearly degenerate coupled
doublet in the yrast band of $^{47}$V \cite{Brandolini01}.
Calculations reproduce well the relative sequence of the yrast
states. A similar scheme is predicted for $^{47}$Cr, as shown in
Fig. \ref{fig5}.

\begin{figure}
\includegraphics[width=0.48\textwidth]{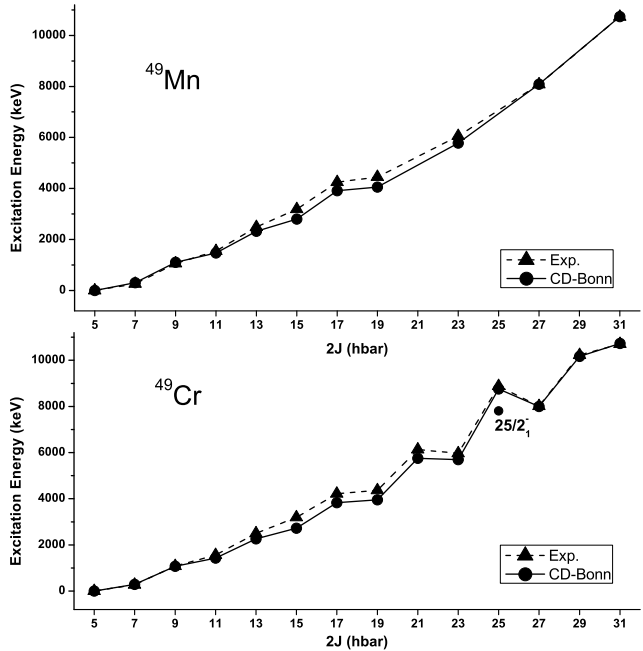}
\caption{\label{fig6} Same as Fig. \ref{fig1} but for $^{49}$Cr and
$^{49}$Mn. Experimental data are taken from Ref. \cite{Oleary97}.}
\end{figure}

\begin{figure}
\includegraphics[width=0.5\textwidth]{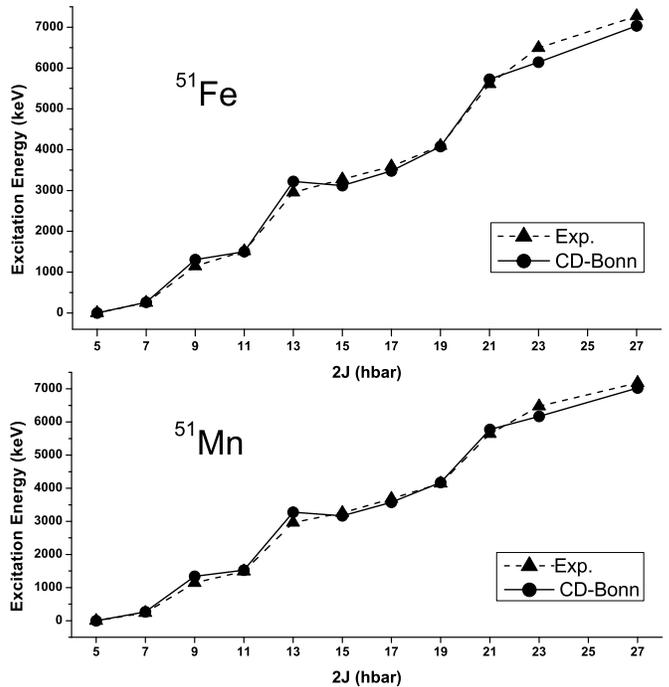}
\caption{\label{fig7} Same as Fig. \ref{fig1} but for $^{51}$Fe and
$^{51}$Mn. Experimental data are taken from Ref. \cite{Bentley00}.}
\end{figure}

Calculated results for mirror pairs $^{49}$Cr-$^{49}$Mn and
$^{51}$Mn-$^{5}$Fe are shown in Fig. \ref{fig6} and \ref{fig7},
respectively. Calculations are done in truncated $fp$ model spaces.
These two pairs are the cross-conjugate partners of the $A=47$ and
$A=45$ nuclei, respectively. For nucleus $^{49}$Cr, one $25/2^-$
state with an excitation energy of 8879 keV was observed in Ref.
\cite{Oleary97}. The calculated level most close to the observed
state locates at 8758 keV. In our calculations, four $25/2^-$ states
are generated around 8879 keV, with the lowest one lying at 7812
keV. The proposed yrast state is more close to the experimental
result of 8334 keV given in Ref. \cite{Brandolini01}. However,
confusion still exists concerning the calculated B(E2) for the
$25/2_1^-\rightarrow 21/2_1^-$ transition when comparing with the
corresponding experimental result, which will be discussed below.

Our calculations in the $fp$ shell generate well the negative-parity
bands for the odd-A nuclei. Positive-parity states can be
approximated by assuming a hole in the $0d_{3/2}$ orbit. In nuclei
$^{45}$Ti and $^{45}$V, the positive-parity collective bands start
at 328 keV and 386 keV, respectively. To give a good description of
the positive-parity states, large-scale calculation in full $sdpf$
shell is needed \cite{Brandolini05}. However, it is beyond the scope
of this paper.

\section{ISB effects in MDE and MED}

As mentioned above, the discrepancy between the theoretical MDE of
mirror nuclei and corresponding experimental data is a long-standing
problem in nuclear physics \cite{Shlomo78,ns,Suzuki92}. In Ref.
\cite{Brown00}, Brown {\it et al.} investigated the role played by
CSB in addressing the NS anomaly in a self-consistent Hartree-Fock
method by adding a CSB term to the $s$-wave part of the Skyrme
interaction. The crucial importance of introducing CSB in partial
waves with $L>0$ for the full explanation of the NS anomaly have
been further studied by M\"uther {\it et al.} within the
Bruckner-Hartree-Fock calculations of nuclear matter
\cite{Muther99}. The anomaly has also been investigated by Tsushima
{\it et al.} \cite{Tsushima99} at the quark level using the
so-called quark-meson coupling (QMC) model in which the
mass-difference between the \textit{up} and \textit{down} quark is
taken into account. QMC can be seen as an extended model of the
relativistic mean-field theory with quark mass difference entering
to account for the short-range CSB \cite{Miller90}.

%%%%%%%%%%%%%%%%%%%%%%% table 1 &&&&&&&&&&&&&&&&&&&&&&&&
\begin{table}
\centering \caption{$0f_{7/2}$-$0f_{7/2}$ monopole interactions and
the monopole centroids. The $T=1$ KB3 matrix elements are also shown
for a comparison.}\label{table1}
\begin{ruledtabular}
\begin{tabular}{ ccccc cc }
$j$ & 0 & 2 & 4 & 6  & centroid\\
\hline
$\pi-\pi$&-1.22&-0.20&0.58 &  0.86&0.51\\
$\nu-\nu$&-1.54&-0.39& 0.40&0.69& 0.32 \\
KB3&-1.92&-1.09& -0.19&0.18 &-0.24\\
\end{tabular}
\end{ruledtabular}
\end{table}

%%%%%%%%%%%%%%%%%%%%%%% table 2 &&&&&&&&&&&&&&&&&&&&&&&&
\begin{table}
\centering \caption{ISB contribution (in MeV) to MDE. See details in
the text.}\label{mde}
\begin{ruledtabular}
\begin{tabular}{ ccccc cc }
Mirror pair& A=45 & A=47 & A=49 & A=51   \\
\hline
T=1/2&0.359&0.604&0.658 &  0.936\\
T=3/2&1.269& 1.586& 2.197&2.579  \\
\end{tabular}
\end{ruledtabular}
\end{table}

In isospin-conserving shell-model calculations, the binding energies
of mirror nuclei are identical since the Coulomb field is not taken
into account. In this work, the Coulomb force and charge-dependence
in strong force are embedded at the two-body level, as shown in
Table \ref{table1}. The two-body ISB interaction leads to more
binding energies for the neutron-rich side of the mirror nuclei. The
binding-energy differences of $T=1/2$ mirror pairs are of the order
of a few hundreds keV. Table \ref{mde} shows results for $T=1/2$ and
$T=3/2$ mirror pairs which is defined as
\begin{equation}
\Delta E_{ISB}=E(Z_<,\text{g.s.})-E(Z_{>},\text{g.s.}),
\end{equation}
where $Z_{<}~(Z_>)$ denotes neutron (proton)-rich side of mirror
nuclei. The energy differences reflect contribution from ISB to MDE.
What is still absent in the present Hamiltonian is the proper
evaluation of Coulomb effect on the SPE. The study of Coulomb effect
on MDE in the context of the shell model has been shown, i.e., in
Ref~\cite{Ormand97}. Further investigations of the problem would be
done in the future to see whether the charge-dependent strong force
can be totally responsible for the NS anomaly.

%%%%%%%%%%%%%%%%   fig 8 %%%%%%%%%%%%%%%%%%%%%%%%%%%%%%%%
\begin{figure*}
\includegraphics[scale=0.81]{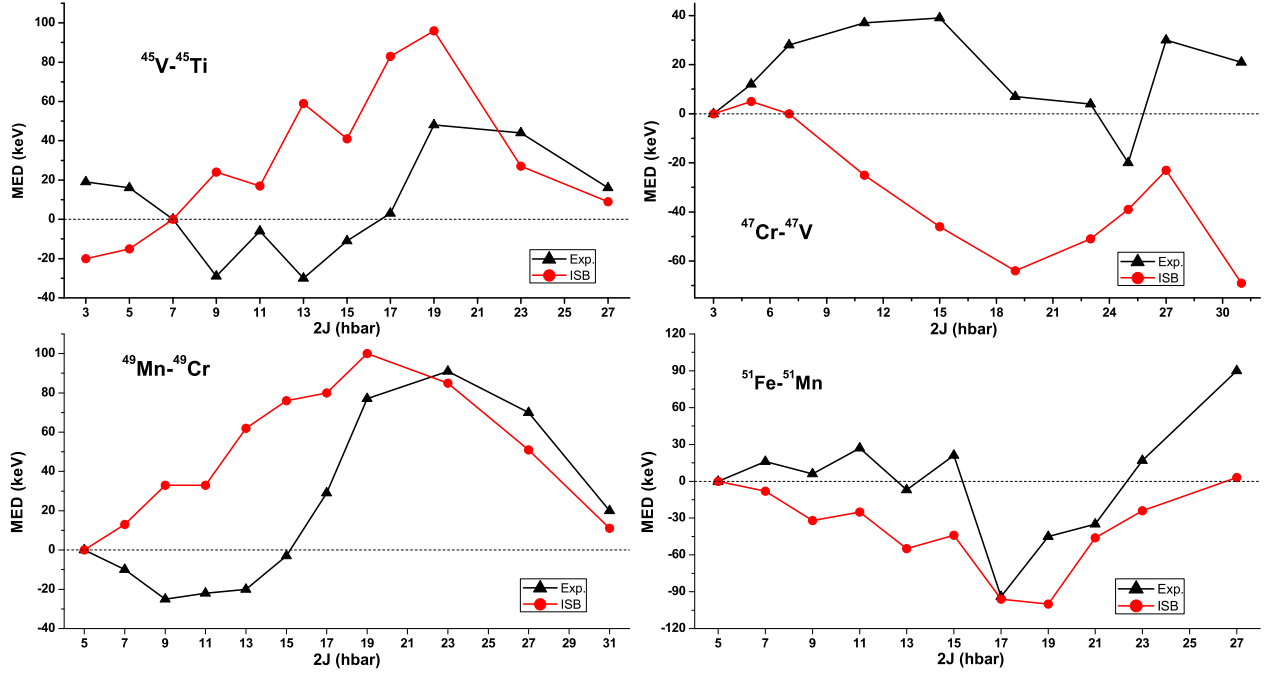}
\caption{\label{med} $\Delta V_M$ contribution (denoted as ISB) to
MED for $T=1/2$ mirror pairs $^{45}$Ti-$^{45}$V, $^{47}$V-$^{47}$Cr,
$^{49}$Cr-$^{49}$Mn and $^{51}$Mn-$^{51}$Fe. Experimental data are
taken from Ref.
\cite{Bentley00,Bentley06,Bentley98,Oleary97,Tonev02}. See the text
for details.}
\end{figure*}

In medium-mass nuclei, observed MED are very small (usually of the
order of 10-100 keV) \cite{Zuker02}. It provides a special ground to
deduce the effect of charge-dependent strong force. In Ref.
\cite{Zuker02}, contribution from CSB are approximated by an
additional $J=2$ pairing term evaluated phenomenally from the MED
and triplet energy difference in the A=42 isospin triplet.
Contributions from the Coulomb field and the above CSB term were
collected and compared with the measured MED in $^{45}$V and
$^{45}$Ti by Bentley {\it et al.} \cite{Bentley06}. However, the
prediction power of the model is very poor \cite{Bentley06},
indicating that a more precise treatment of the charge-dependence in
strong force is needed.

In the present work, contributions from the two-body Coulomb force
and the charge-dependent strong force can be calculated exactly from
the calculated energy differences of analogous states in mirror
nuclei. We simply decompose the contributions to total MED into two
parts,
\begin{equation}
\text{MED}_J=\Delta\langle U_s\rangle_J+\Delta (V_{M})_J,
\end{equation}
where $\Delta\langle U_s\rangle_J$ denotes contribution from the
Coulomb shifts of proton SPE. $\Delta (V_{M})_J$ is given as
\begin{equation}
\Delta (V_{M})_J=E_{\text{cal.}}(Z_>)_J-E_{\text{cal.}}(Z_<)_J,
\end{equation}
where $E_{\text{cal.}}$ is calculated excitation energy of the state
with angular momentum $J$. Calculations for $\Delta (V_{M})_J$ are
plotted in Fig. \ref{med}.

Configurations for low-lying states in the lower $fp$ shell nuclei
are dominated by the $0f_{7/2}$ orbit since a sizable energy-gap
between the $0f_{7/2}$ and other orbits ($1p_{3/2}, 0f_{5/2},
1p_{1/2}$) exists. Contributions from the monopole part of the
Coulomb force, however, are sensitive to the $1p_{3/2}$ occupancy
and are expected to be enhanced at the nuclear surface
\cite{Bulgac99}. In recent studies, explanations of the evolution
behavior of MED with spin have been focus on particle alignment and
nuclear shape changes \cite{Bentley00,Oleary97,Bentley98}. These
arguments are mainly based on the fact that the MED tend to vanish
at band termination states where maximum possible alignments of all
valence particles are expected.

In the MED of the $A=51$ mirror pair, two sharp changes exist,
leading to two peaks of about 100 keV at the $J=17/2^-$ state and
the band-terminated $J=27/2^-$ state. Calculations with empirical
Coulomb matrix elements reproduced the two peaks but over-estimated
results for the $J=17/2^-$ state \cite{Bentley00}. Further
investigations by the authors showed that a sharp alignment of a
$0f_{7/2}$ proton pair exist at the $17/2^-$ state in $^{51}$Fe
\cite{Bentley00}. In our calculations, the ISB contribution $\Delta
(V_{M})_J$ reproduced the peak at the $J=17/2^-$ state and vanished
at the termination state.

Also interesting is the staggering in the MED of yrast bands in
mirror pair $^{45}$Ti and $^{45}$V and the cross-conjugate partners
$^{51}$Fe and $^{51}$Mn. In Ref. \cite{Bentley00}, it was
tentatively explained as the existence of two bands with different
signatures in the yrast band. Calculated $\Delta (V_{M})_J$
characterized the staggering.

\section{Transition properties in mirror nuclei}

Electromagnetic (EM) and Gamow-Teller (GT) analogous transitions in
mirror nuclei can provide other detailed information on the
properties of nuclear structures and transition operators
\cite{Ekman04,Fujita04,Smirnova03}. Different transition strengths
have already been identified in the yrast cascades of $^{47}$V and
$^{47}$Cr, with tentative work trying to extract information on
differences in corresponding wave functions by Tonev {\it et al.}
\cite{Tonev02}. In the following, calculated strengths for EM and GT
analogous transitions will be given, with the analyse of existing
asymmetries.

%%%%%%%%%%%%%%%%%%%% table III & IV %%%%%%%%%%%%%%%%%%%%%%%%%%%%%%
\begin{table*}
\centering \caption{Calculated B(E2) (in e$^2$fm$^4$) in mirror
nuclei $^{45}$Ti and $^{45}$V. Effective charges with e$_{p}$=1.5e
and e$_n$=0.5e are used.}\label{table2}
\begin{ruledtabular} \begin{tabular}{cccccccccc}
&\multicolumn{4}{c}{$^{45}$Ti}~~~  &&\multicolumn{4}{c}{$^{45}$V}\\
\cline{2-5}\cline{7-10}
$J_i\rightarrow J_f$ &HO&WS&SKcsb&KB3&~~~&HO&WS&SKcsb&KB3\\
\hline
$11/2^-_1\rightarrow 7/2^-_1$  &124&124&105& 132 && 154&158&134&157\\
$15/2^-_1\rightarrow 11/2^-_1$  &139&139& 118&144&& 167 &171&146&165\\
$19/2^-_1\rightarrow 15/2^-_1$&96&95&80&94&&110&113&95&98\\
$23/2^-_1\rightarrow 19/2^-_1$&65&64&54&70&&119&122&104&118\\
$27/2^-_1\rightarrow 23/2^-_1$&48&47&40&50&&87&90&76&83\\
$9/2^-_1\rightarrow 5/2^-_1$ &  82 &86&71&94&& 97&101&85&104 \\
$13/2^-_1\rightarrow 9/2^-_1$&151&160&131&158&&175&185&155&191\\
$17/2^-_1\rightarrow 13/2^-_1$&100&101&84&108&&130&135&113&130\\
$9/2^-_1\rightarrow 7/2^-_1$ &  48 &51&41&83&& 41&45& 36&77\\
$13/2^-_1\rightarrow 11/2^-_1$ &  33&35&29 &44&& 28&30&25&31 \\
$15/2^-_1\rightarrow 13/2^-_1$ & 16 &17&14&22&& 30&33& 26&27\\
$17/2^-_1\rightarrow 15/2^-_1$ &  14 &14&12&22&& 9 &10&8&11\\
$19/2^-_1\rightarrow 17/2^-_1$ &  17 &17&15&17&& 38 &40&34&30\\
\end{tabular}
\end{ruledtabular}
\end{table*}

\begin{table}
\centering \caption{Theoretical B(M1) (in $\mu^2_N$) and B(GT) in
mirror pair $^{45}$Ti and $^{45}$V. Free nucleon g-factors
[g$_{s}$(proton)=5.586, g$_s$(neutron)=-3.826, g$_{l}$(proton)=1 and
g$_{l}$(neutron)=0] and free axial-vector constant ($g_A=1.26$) are
used.}\label{a45m1}
\begin{ruledtabular}
\begin{tabular}{cccccc}
& \multicolumn{2}{c}{B(M1)}&~~ &\multicolumn{2}{c}{B(GT)}\\
\cline{2-3}\cline{5-6}
$J_i\rightarrow J_f$&$^{45}$Ti&$^{45}$V&&$\beta^+$&CE\\
\hline
$9/2^-_1\rightarrow 7/2^-_1$    & 0.40  & 0.42   &~~ &  0.12&0.13 \\
$13/2^-_1\rightarrow 11/2^-_1$   &  0.21 & 0.25  &~~&  0.08 &0.09\\
$15/2^-_1\rightarrow 13/2^-_1$   & 0.87  &  0.83   &~~&  0.30&0.32 \\
$17/2^-_1\rightarrow 15/2^-_1$   & 0.11  &  0.13  &~~&  0.04&0.05\\
$19/2^-_1\rightarrow 17/2^-_1$   & 0.15  &  0.16  &~~ &  0.53&0.53\\
\end{tabular}
\end{ruledtabular}
\end{table}
%%%%%%%%%%%%%%%%%%%%%%%%%%%%%%%%%%%%%%%%%%%%%%%%%%%%%%%%%%%%%%%%%%%

Table \ref{table2} shows calculated strengths for electrical
quadrupole (E2) analogous transitions in $^{45}$Ti and $^{45}$V.
Effective charges for neutrons and protons have been renormalized to
e$_{n}$=0.5e and e$_p$=1.5e, respectively, to account for the
core-polarization effect. The wave functions of the shell model are
calculated with three different bases, i.e., the HO, the Woods-Saxon
(WS) potential and the Skyrme force (SKcsb) by Brown {\it et al.}
\cite{Brown00} in which a charge-symmetry-breaking term has been
added. For comparison, results calculated with the KB3 interaction
and the HO basis (denoted as KB3) are also shown.

%%%%%%%%%%%%%%%%%%%%% table  V & VI & VII %%%%%%%%%%%%%%%%%%%%%%%%%%%%%%%%%%%%
\begin{table}
\centering \caption{Calculated B(E2) (in e$^2$fm$^4$) and B(M1) (in
$\mu^2_N$) for analogous transitions in mirror pair $^{47}$V and
$^{47}$Cr.}\label{table4}
\begin{ruledtabular}  \begin{tabular}{cccccc}
&\multicolumn{2}{c}{B(E2)}&~~ &\multicolumn{2}{c}{B(M1)}\\
\cline{2-3}\cline{5-6}
$J_i\rightarrow J_f$~~~~& ~~~$^{47}$V~~~&~~~$^{47}$Cr~~~&~~~&~~~$^{47}$V~~~~&~~$^{47}$Cr~~~\\
\hline
$7/2^-_1\rightarrow 3/2^-_1$    & 142  & 172   &~~ &   &  \\
$11/2^-_1\rightarrow 7/2^-_1$   &  242 & 314  &~~&    & \\
$15/2^-_1\rightarrow 11/2^-_1$   & 249  &  326   &~~&   &  \\
$19/2^-_1\rightarrow 15/2^-_1$   &203  &  261  &~~&   & \\
$23/2^-_1\rightarrow 19/2^-_1$   & 182  &  221  &~~ &    & \\
$27/2^-_1\rightarrow 23/2^-_1$   &118  &  141  &~~&   & \\
$31/2^-_1\rightarrow 27/2^-_1$   & 67  &  96  &~~ &   & \\
$9/2^-_1\rightarrow 5/2^-_1$    & 182  & 248   &~~ &   &  \\
$13/2^-_1\rightarrow 9/2^-_1$    & 220  &  291   &~~ & &  \\
$17/2^-_1\rightarrow 13/2^-_1$    & 218  &   256  &~~ &   &  \\
$21/2^-_1\rightarrow 17/2^-_1$    & 185  &  225   &~~ &   &  \\
$25/2^-_1\rightarrow 21/2^-_1$    & 121  &  141  &~~ &   &  \\
$29/2^-_1\rightarrow 25/2^-_1$    & 1  &   1  &~~ &   &  \\
$5/2^-_1\rightarrow 3/2^-_1$   &314  &  393  &~~& 0.15  &0.13 \\
$7/2^-_1\rightarrow 5/2^-_1$   & 251 &  282  &~~& 0.29  &0.26\\
$9/2^-_1\rightarrow 7/2^-_1$   & 101 &  137  &~~& 0.11  &0.10\\
$13/2^-_1\rightarrow 11/2^-_1$   & 43  &  60  &~~&  0.07 & 0.06\\
$15/2^-_1\rightarrow 13/2^-_1$   & 74 &  81 &~~& 0.70  & 0.62\\
$17/2^-_1\rightarrow 15/2^-_1$   & 17 &  25   &~~& 0.03  &0.03 \\
$17/2^-_1\rightarrow 19/2^-_1$   & 37 &  49   &~~& 0.70  &1.0 \\
$21/2^-_1\rightarrow 19/2^-_1$   & 2.4 & 8.3    &~~& 0.11  &0.12 \\
$21/2^-_1\rightarrow 23/2^-_1$   & 2.9 &  5  &~~& 1.6  & 1.5\\
$25/2^-_1\rightarrow 23/2^-_1$   & 0.06 & 2.7    &~~& 0.06  & 0.07\\
$27/2^-_1\rightarrow 25/2^-_1$   & 4.1 &  6.2 &~~& 2.2  &2.1 \\
$29/2^-_1\rightarrow 31/2^-_1$   & 0.24 &  0.10   &~~& 0.27$\times10^{-2}$  &0.18$\times10^{-2}$ \\
$29/2^-_1\rightarrow 27/2^-_1$   & 5.4 &   18  &~~&  0.10$\times10^{-1}$  & 0.15$\times10^{-1}$\\
\end{tabular}
\end{ruledtabular}
\end{table}

\begin{table}
\centering \caption{Theoretical B(E2) (in e$^2$fm$^4$) and B(M1) (in
$\mu^2_N$) for analogous transitions in mirror pair $^{49}$Cr and
$^{49}$Mn.}\label{table5}
\begin{ruledtabular}  \begin{tabular}{cccccc}
& \multicolumn{2}{c}{B(E2)}&~~ &\multicolumn{2}{c}{B(M1)}\\
\cline{2-3}\cline{5-6}
$J_i\rightarrow J_f$& $^{49}$Cr&$^{49}$Mn&&$^{49}$Cr&$^{49}$Mn\\
\hline
$9/2^-_1\rightarrow 5/2^-_1$    & 110  & 109   &~~ &   &  \\
$13/2^-_1\rightarrow 9/2^-_1$    & 223  &  225   &~~ & &  \\
$17/2^-_1\rightarrow 13/2^-_1$    & 191  &  214  &~~ &   &  \\
$21/2^-_1\rightarrow 17/2^-_1$    & 190 &  246   &~~ &   &  \\
$25/2^-_1\rightarrow 21/2^-_1$    & 148  &  231   &~~ &   &  \\
$29/2^-_1\rightarrow 25/2^-_1$    & 165  &  224   &~~ &   &  \\
$11/2^-_1\rightarrow 7/2^-_1$   &  195 & 204  &~~&    & \\
$15/2^-_1\rightarrow 11/2^-_1$   & 228  &  230   &~~&   &  \\
$19/2^-_1\rightarrow 15/2^-_1$   & 211  &  223  &~~&   & \\
$23/2^-_1\rightarrow 19/2^-_1$   & 191  &  224  &~~ &    & \\
$27/2^-_1\rightarrow 23/2^-_1$   &131  &  159  &~~&   & \\
$31/2^-_1\rightarrow 27/2^-_1$   & 51  &  55   &~~&   &  \\
$7/2^-_1\rightarrow 5/2^-_1$   & 382 &  408  &~~& 0.24  &0.26\\
$9/2^-_1\rightarrow 7/2^-_1$   & 307 &  299   &~~& 0.57 &0.58  \\
$11/2^-_1\rightarrow 9/2^-_1$   & 234 &  247  &~~&  0.65 &0.66\\
$13/2^-_1\rightarrow 11/2^-_1$   & 165  &  162  &~~& 0.70  & 0.73\\
$15/2^-_1\rightarrow 13/2^-_1$   & 106 &  130 &~~&  0.88 & 0.93\\
$17/2^-_1\rightarrow 15/2^-_1$   & 67 &  89   &~~&  0.29 &0.33 \\
$19/2^-_1\rightarrow 17/2^-_1$   & 92 &  131   &~~&  0.42 &0.53 \\
$21/2^-_1\rightarrow 19/2^-_1$   &  38 &  72  &~~&  0.48$\times10^{-1}$ & 0.63$\times10^{-1}$ \\
$25/2^-_1\rightarrow 23/2^-_1$   & 29 &  61  &~~& 0.30$\times10^{-3} $ & 0.68$\times10^{-5} $\\
$29/2^-_1\rightarrow 27/2^-_1$   & 22 &  45  &~~& 0.47$\times10^{-2}$  & 0.64$\times10^{-2}$\\
$31/2^-_1\rightarrow 29/2^-_1$   & 3.0 &  10  &~~&  0.58  & 0.75\\
\end{tabular}
\end{ruledtabular}
\end{table}

\begin{table}
\centering \caption{Calculated B(E2) (in e$^2$fm$^4$) and B(M1) (in
$\mu^2_N$) for analogous transitions in mirror pair $^{51}$Mn and
$^{51}$Fe.}\label{table6}
\begin{ruledtabular}  \begin{tabular}{cccccc}
& \multicolumn{2}{c}{B(E2)}&~~ &\multicolumn{2}{c}{B(M1)}\\
\cline{2-3}\cline{5-6}
$J_i\rightarrow J_f$& $^{51}$Mn&$^{51}$Fe&&$^{51}$Mn&$^{51}$Fe\\
\hline
$9/2^-_1\rightarrow 5/2^-_1$    & 81  & 95   &~~ &   &  \\
$13/2^-_1\rightarrow 9/2^-_1$    & 155  &  183   &~~ & &  \\
$17/2^-_1\rightarrow 13/2^-_1$    & 1  &   1  &~~ &   &  \\
$21/2^-_1\rightarrow 17/2^-_1$    & 36  &  18   &~~ &   &  \\
$11/2^-_1\rightarrow 7/2^-_1$   &  156 & 179  &~~&    & \\
$15/2^-_1\rightarrow 11/2^-_1$   & 189  &  232   &~~&   &  \\
$19/2^-_1\rightarrow 15/2^-_1$   & 54  &  63  &~~&   & \\
$23/2^-_1\rightarrow 19/2^-_1$   & 81  &  52  &~~ &    & \\
$27/2^-_1\rightarrow 23/2^-_1$   &69  &  51  &~~&   & \\
$7/2^-_1\rightarrow 5/2^-_1$   & 294 &  205  &~~& 0.23  &0.22\\
$9/2^-_1\rightarrow 7/2^-_1$   & 183 &  197   &~~& 0.15  &0.14 \\
$11/2^-_1\rightarrow 9/2^-_1$   & 178 &  153  &~~& 0.48  &0.45\\
$13/2^-_1\rightarrow 11/2^-_1$   & 63  &  67  &~~&  0.08 & 0.07\\
$15/2^-_1\rightarrow 13/2^-_1$   & 89 &  74 &~~& 0.71  & 0.68\\
$17/2^-_1\rightarrow 15/2^-_1$   & 0.01 &  1.3   &~~& 0.38$\times10^{-2}$  &0.40$\times10^{-2}$ \\
$19/2^-_1\rightarrow 17/2^-_1$   & 104 &  51   &~~& 0.77  &0.83 \\
$21/2^-_1\rightarrow 19/2^-_1$   & 86 & 47   &~~& 0.83  &0.85 \\
$23/2^-_1\rightarrow 21/2^-_1$   & 43 &  37  &~~& 1.65  & 1.68\\
\end{tabular}
\end{ruledtabular}
\end{table}
%%%%%%%%%%%%%%%%%%%%%%%%%%%%%%%%%%%%%%%%%%%%%%%%%%%%%%%%%%%%%

\begin{figure}
\includegraphics[scale=0.35]{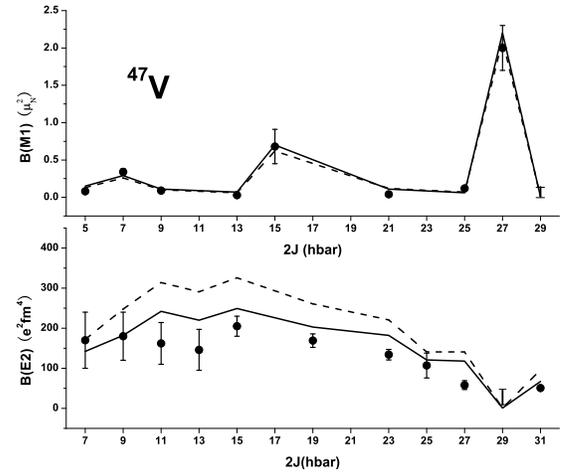}
\caption{\label{fig9}Comparisons between experimental
\cite{Brandolini01} and calculated (solid line) $B(M1)_{J\rightarrow
J-1}(\mu_N^2)$ and $B(E2)_{J\rightarrow J-2}$(e$^2$fm$^4$) in the
yrast band of $^{47}$V. The dashed line gives corresponding
theoretical results for $^{47}$Cr.}
\end{figure}

\begin{figure}
\includegraphics[scale=0.35]{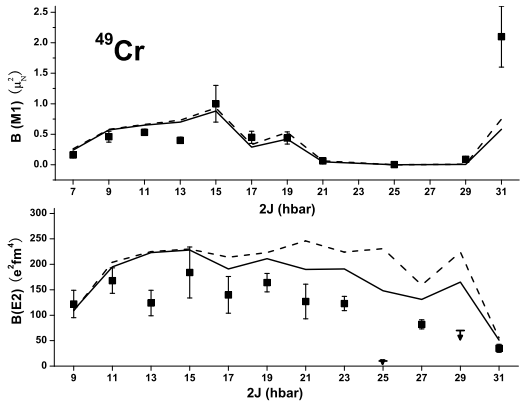}
\caption{\label{fig10}Same as Fig. \ref{fig9} but for $^{49}$Cr.
Experimental data are taken from \cite{Brandolini01}. The dashed
line gives corresponding theoretical strengths in $^{49}$Mn.}
\end{figure}

B(E2) strengths derived from the three bases are similar to each
other. In the following, only results with the HO basis are given
for simplicity. Calculated B(E2) values for mirror pairs
$^{47}$V-$^{47}$Cr, $^{49}$Cr-$^{49}$Mn, and $^{51}$Mn-$^{51}$Fe are
given in Table \ref{table4},\ref{table5}, and \ref{table6},
respectively. Comparisons with available experimental data are
ploted in Fig. \ref{fig9} and \ref{fig10}. It can be seen that
overall agreements are excellent. Our calculations reproduce the
observed staggering pattern in the EM transitions of $^{47}$V and
$^{49}$Cr. The largest discrepancy appears at the $25/2^-_1$ state
in $^{49}$Cr at which rather small E2 strength has been observed
\cite{Brandolini01}. A large B(E2) value, however, is expected for
the yrast decay in our calculations. Confusion still exist
concerning the position of the $25/2^-$ yrast state
\cite{Oleary97,Brandolini01}. More experimental and theoretical
efforts may clear the picture.

The M1 transition is relatively insensitive to the radial property
of the wave function. The transition operator is given by
\begin{equation}\label{m1}
\hat{O}(M1) =
\left[\sum_{n=1}^Z(g_l^{\pi}\mathbf{l}_n+g_s^{\pi}\mathbf{s}_n)
+\sum_{n=1}^N(g_l^{\nu}\mathbf{l}_n+g_s^{\nu}\mathbf{s}_n)\right]\mu_N, \\
\end{equation}
where $\mu_N$ is the nuclear magneton and $g_{l(s)}$ the orbital
(spin) gyromagnetic factor. Eq. (\ref{m1}) can be rewritten as
\begin{eqnarray}\label{m2}
\nonumber \hat{O}(M1)&=&  \Bigg[\sum_{n=1}^A\left(\frac{g_l^{\pi}+
g_l^{\nu}}{2}\mathbf{l}_n+\frac{g_s^{\pi}+g_s^{\nu}}{2}
\frac{\text{\boldmath{$\sigma$}}_n}{2}\right)\\
&-&\left(\frac{g_l^{\pi}-g_l^{\nu}}{2}\mathbf{l}_n+\frac{g_s^{\pi}-g_s^{\nu}}{2}
\frac{\text{\boldmath{$\sigma$}}_n}{2}\right)\tau_z(n)\Bigg]\mu_N,
\end{eqnarray}
with which M1 transition strengths can be separated into two parts:
the isoscalar and isovector term. It can be seen that the M1
transition strength is dominated by the isovector spin
($\sigma\tau_z$) term with coupling constant of
$g_s^{IV}=(g_s^{\pi}-g_s^{\nu})/2$=4.706. Contributions from orbital
terms are expected to be enhanced when nuclear deformation effects
manifest.

Strengths for analogous M1 transitions are calculated with the free
gyromagnetic factors of g$_{s}^{\pi}$=5.586, g$_s^{\nu}$=-3.826,
g$_{l}^{\pi}$=1 and g$_{l}^{\nu}$=0. Results for the $A=45$, 47, 49
and 51 mirror pairs are given in Table \ref{a45m1}, \ref{table4},
\ref{table5} and \ref{table6}, respectively. Fig. \ref{fig9} and
\ref{fig10} show comparisons with experiments. Our calculations
reproduced well the experimental strengths except the strong M1
strength observed at the band-terminated state in $^{49}$Cr.

Weak processes in atomic nuclei can be separated into two kinds, the
Fermi (isoscalar and spin-unflip) and Gamow-Teller (isovector and
spin-flip) transitions. The Lorentz covariant hadronic current
related to the GT transition can be written as
\begin{equation}\label{avector}
A_{\mu}=i\overline{\psi}_p[g_A\gamma_{\mu}\gamma_5+\frac{g_T}{2M}\sigma_{\mu\nu}
\gamma_5k_{\nu}+ig_Pk_{\mu}\gamma_5]\psi_n,
\end{equation}
where $k_{\mu}$ is the transferred momentum, $M$ the mass of the
nucleon and $\psi_p~(\psi_n)$ the proton (neutron) field operator.
Included in the bracket are the axial-vector, induced tensor and
induced pseudoscalar term, with $g_A,~g_T$ and $g_P$ the
corresponding coupling constants. In atomic nuclei, processes
governed by the hadronic current are dominated by the isovector
term. The isovector GT transition operator is given as
\begin{equation}
\hat{O}(\text{GT}^{\pm})=\frac{1}{2}g_A\sum_{n=1}^A\bm{\sigma}_n\tau_{\pm}(k),
\end{equation}
where $\bm{\sigma}$ and $\tau$ are spin and isospin operators,
respectively.

\begin{figure}
\includegraphics[scale=0.42]{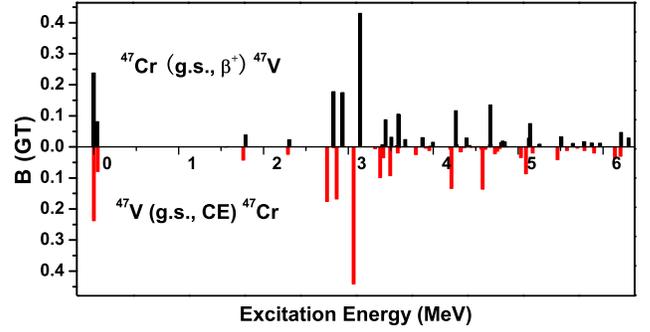}
\caption{\label{fig11} Ground-state GT transition strengths for the
$\beta^+$-decay of $^{47}$Cr and the CE reaction of $^{47}$V. Free
axial-vector ($g_A=1.26$) are used.}
\end{figure}

\begin{figure}
\includegraphics[scale=0.42]{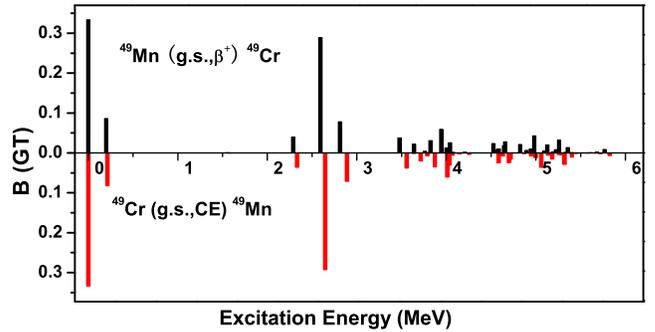}
\caption{Same as Fig. \ref{fig11} but for the $A=49$
pair.\label{fig12}}
\end{figure}

\begin{figure}
\includegraphics[scale=0.42]{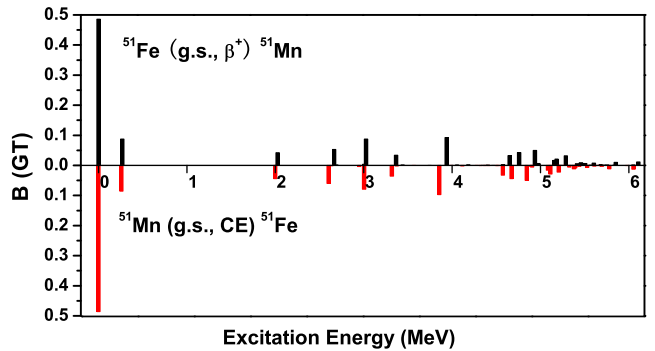}
\caption{Same as Fig. \ref{fig11} but for the $A=51$
pair.\label{fig13}}
\end{figure}

Experimentally, B(GT) is accessible through the study of
$\beta$-decay and the conjugate charge-exchange (CE) reaction, such
as $(p,n)$ reaction. All nuclei discussed above have $\beta^+$ decay
mode. The distributions of the reduced GT transition strengths for
ground-state $\beta^+$-decay of $^{47}$Cr, $^{49}$Mn, and $^{51}$Fe
are plotted in Fig. \ref{fig11}, \ref{fig12}, and \ref{fig13},
respectively. In the lower part of the figures, we give the B(GT)
values of conjugate CE processes as a comparison. If isospin
symmetry is exactly conserved, the two strength distributions should
be identical. The asymmetry deduced from the strengths,
\begin{equation}
\delta^{ISB}=\frac{B(\text{GT;~CE})}{B(\text{GT};~\beta^+)}-1,
\end{equation}
reflects the ISB effect. Another possible origin of the observed
asymmetry come from the non-zero contribution of the induced-tensor
term (commonly referred to as second-class current) in Eq.
(\ref{avector}). The underlying electroweak theory does not put any
limit on the value of $g_T$. Investigations for the possible
existence of induced-tensor current in weak processes is
longstanding \cite{Smirnova03}. In the past a few decades,
considerable attentions have been paid to light nuclei where large
asymmetry has been identified in mirror nuclei GT transitions (See
Ref. \cite{Smirnova03} for a review). The dominant origin is the
strong Coulomb effect which, however, can lead to large
uncertainties. One notable property of \textit{fp} shell nuclei is
the relatively small influence from the Coulomb field. More
extensive studies concerning the \textit{fp} nuclei would to very
helpful.

Free axial-vector factor with $g_A=1.26$ are used in the
calculations. For practical applications, a renormalization factor,
about 0.744 in $fp$ shell, has to be introduced to account for the
quenching effects induced by core-polarization and subnuclear
freedoms \cite{Caurier05}.

Contributions from orbital and spin terms to M1 transition strengths
in $T=1/2$ nuclei can be separated since both isovector M1 and GT
decay processes in nuclei are dominated by a $\sigma\tau$-type
operator \cite{Fujita04}. If isospin symmetry is exact, the final
states of the M1 transition and the corresponding GT transition
differ only in their $\tau_z$ quantum number. Relative strengths
between GT and isovector spin-flip M1 strengths can be written as
\begin{equation}
\frac{B(GT^{\pm})}{B(\sigma\tau_z)}=\frac{8\pi}{3}\left(\frac{g_A}
{g_s^{IS}}\right)^2\frac{\langle
T_i,T_{iz},1,\pm1|T_f,T_{fz}\rangle^2} {\langle
T_i,T_{iz},1,0|T_f,T_{fz}\rangle^2},
\end{equation}
by which a more deep insight into the nuclear structure can be
deduced. It also has direct astrophysics applications
\cite{Kawabata04}. Some similar works have been done in $p$ and $sd$
shell, as shown in Ref. \cite{Fujita04}. ISB effects on the relation
can be evaluated through the calculations of direct overlaps between
the analogous final states. We caulcated M1 and GT transitions
connecting analogous states in the $T=1/2$ pairs. Theoretical
results for B(M1) and B(GT) in mirror pair $^{45}$Ti and $^{45}$V
are shown in Table \ref{a45m1}.

\section{Summary}
$T=1/2$ mirror nuclei in the lower $fp$ shell have been studied by
shell-model configuration-mixing calculations based on a microscopic
effective Hamiltonian. The effective interaction is derived from the
high-precision CD-Bonn \textit{NN} potential. Calculated level
schemes agree well with experimental observations till the
$0f_{7/2}$ band termination. The largest discrepancy seen at the
$25/2^-$ state in $^{49}$Cr is also discussed. Electromagnetic and
Gamow-Teller transition strengths for analogous transitions in the
mirror pairs are presented where experimental observations are still
insufficient.

The isospin-nonconserving effective interaction enables the
investigation of asymmetries existing in nuclear structures and
transitions. We calculate ISB contributions to MDE and MED of the
mirror pairs. An important role played the term is identified. The
discrepancies exist in analogous transitions ara analyzed which can
be used to deduce properties of nuclear structures and decay
operators.

\begin{acknowledgments}
This work has been supported by the Natural Science Foundations of
China under Grant Nos. 10525520 and 10475002, the Key Grant Project
(Grant No. 305001) of Ministry of Education of China. We also thank
the PKU computer Center where numerical calculations have been done.
\end{acknowledgments}

\end{document}